\documentclass[preprint]{revtex4}%
\usepackage{amssymb}
\usepackage{amsmath}
\usepackage{amsfonts}
\usepackage{graphicx}
\usepackage{epstopdf}

\begin{document}

\title{Dark/Visible Parallel Universes and Big Bang Nucleosynthesis}


\author{C. A. Bertulani$^1$\footnote{email: carlos.bertulani@tamuc.edu}, T. Frederico$^2$, J. Fuqua$^1$,   M. S. Hussein$^3$,  O. Oliveira$^{2,4}$,
and W. de Paula$^2$ }
\affiliation{$^1$Department of Physics and Astronomy, Texas A\&M University-Commerce,
Commerce TX 75429, USA\\
                                     $^2$Departamento de F\'{\i}sica, Instituto Tecnol\'ogico de Aeron\'autica, DCTA
                               12.228-900, S\~ao Jos\'e dos Campos, SP, Brazil \\
$^3$ Instituto de F\'{\i}sica, Universidade de S\~ao Paulo, Caixa Postal 66318,
05314-970 S\~ao Paulo, SP, Brazil\\
                               $^4$Departamento de F\'{\i}sica, Universidade de Coimbra, 3004-516 Coimbra, Portugal 
                               }

\begin{abstract}
We develop a model for visible matter-dark matter interaction based on the exchange of a massive gray boson called herein the Mulato. Our model hinges on the assumption that all known particles in the visible matter have their counterparts in the dark matter. We postulate six families of particles five of which are dark. This leads to the unavoidable postulation of six parallel worlds, the visible one and five invisible worlds.  A close study of big bang nucleosynthesis (BBN), baryon asymmetries, cosmic microwave 
background (CMB) bounds, galaxy 
dynamics,  together with the Standard Model assumptions, help us to set a limit on the mass and width 
of the new gauge boson. Modification of the statistics underlying the kinetic energy distribution of particles during the BBN is also discussed. The changes in reaction rates during the BBN due to a departure from the Debye-Hueckel electron screening model is also investigated.  
\end{abstract}

\maketitle


\section{Dark/Cold Universes and  the Mulato Boson} 

The mass density ratios computed from the Wilkinson Microwave Anisotropy Probe (WMAP) \cite{WMAP1} data
show that the present day dynamics of the Universe is driven essentially by the Dark Energy (DE), 
see e.g. \cite{Li2011}, Indeed, while $\Omega_{DE} = 0.734\pm0.029$ 
for ordinary baryonic matter, i.e. nuclei and electrons, $\Omega_{b} = 0.0449\pm0.0028$ which is around 
5 times smaller than the corresponding value for dark matter (DM), $\Omega_{DM} = 0.222\pm0.026$.

The nature of dark matter (DM) is a fundamental problem in modern physics. 
Dark matter, see e.g. \cite{Feng2010,bertone05,bertone10},
 is a form of matter that does not interact significantly with ordinary baryonic matter.
Experimental evidence for dark matter comes from the anisotropies of CMB and the dynamics of galaxy clusters.
Elementary particle theory offer scenarios
where new particles such as Weakly Interacting Massive Particles (WIMPs),
Sterile Neutrinos, Axions, Supersymmetric Particles, etc., are possible candidates for DM.

A possible scenario for dark matter is the presence of a mirror(s) sector(s) of particles
\cite{Yang,Kobzarev,Pavsic,Foot,Akhmedov} where the mirror sectors are copies of the Standard Model (SM).
The mirror sectors are not necessarily exact copies of the Standard Model, with, e.g. the mirror particles
having different masses and/or couplings than the corresponding SM particles.
Anyway, ordinary and mirror particles 
are weakly coupled. Different mirror models provide different mechanisms for the coupling between 
ordinary matter and DM.

We developed a mirror model which relies on gauging a symmetry which was so far 
not completely explored \cite{Orlando}.
Classifying the fundamental matter fields of the Standard Model according to their electric charge leads, 
quite naturally, to an $SU(3)$ symmetry, which can be made local to give dynamics to the interaction.
The model does not requires an \textit{a priori} number of mirror sectors. However, if the dark sectors
are exact copies of the SM, to explain the relative abundance between ordinary and dark matter, five
dark sectors are required. Note that using the quoted values for $\Omega_{DM}$ and
$\Omega_{b}$ it follows that $\Omega_{DM} / \Omega_{b} =4.94 \pm 0.66$; the error on the ratio
was computed assuming gaussian error propagation. Of course, besides the relative abundance the
model should be made compatible with the known cosmological constraints,
with Big Bang Nucleosynthesis (BBN) and with the experimental bounds on the cross sections for
the interaction with ordinary matter.

The gauge model discussed here explores a  SU(3) symmetry and introduces a 
new Weakly Interacting Massive Gauge boson (WIMG) which couples the different sectors and, in this way,
provides the link between dark and ordinary matter. The WIMG, called herein the ``{\it Mulato}", being a massive boson, leaves unchanged the 
long distance properties of the SM and gravity. The model is compatible with BBN and the recent measurements of the CMB. Further,
the dark sectors associated with multiple universes of dark matter can be made collisionless if the temperature of the dark sectors is sufficiently lower than 
the observed temperature of the visible universe. This difference in the temperature seems to suggest
that the dark sectors are not exact copies of the SM sector.

\section{Mulato Coupling and Mass \label{modelo}} 

At energies much larger than the typical electroweak scale, the SM matter fields behave
like massless particles. It is natural to group the matter fields according to their \cite{Orlando}
electric charge
\begin{eqnarray}
 & &  Q_1 = \left( \begin{array}{c} u \\ c \\ t \end{array} \right) , \quad
  Q_2 = \left( \begin{array}{c} d \\ s \\ b \end{array} \right) , \nonumber \\
  & &
  Q_3 = \left( \begin{array}{c} e \\ \mu \\ \tau \end{array} \right) , \quad
  Q_4 = \left( \begin{array}{c} \nu_e \\ \nu_\mu \\ \nu_\tau \end{array} \right) \, .
  \label{matter}
\end{eqnarray}
In the following, we will also use the notation $\mathcal{Q} = \left\{ Q_1, Q_2, Q_3, Q_4 \right\}$.
We assume that the DM has a similar structure as observed for ordinary matter.
Each DM sector $\mathcal{Q}_s$ has 4 multiplets which mimic 
(\ref{matter}) and each sector has its own copy of the SM. Each sector has its own copy of the SM, with 
the corresponding electroweak sectors bosons coupling only within the sector that they are associated with.
Our gauge mirror model includes the Mulato gauge field $M^a_\mu$, 
the matter fields $Q_{i,s}$, where the new index $s$ distinguishes between the different $N_Q$ sectors
and $i$ qualifies the fermions as in (\ref{matter}).
A real scalar field $\phi^a$ belonging to the adjoint representation of the $SU(3)_Q$ group is introduced
as an effective way to provide a mass to $M^a_\mu$, ensuring that the Mulato interaction is short ranged.
The Lagrangian  for the gauge theory reads
\begin{eqnarray}
 \mathcal{L}   =   - \frac{1}{4} F^a_{\mu\nu} F^{a \, \mu\nu} ~ + ~
 \sum^{N_Q}_{s=1}\sum^4_{i=1} \overline Q_{i,s} \, i \gamma^\mu D_\mu \, Q_{i,s}
 ~ + ~ 
 + ~ \frac{1}{2} \left( D^\mu \phi^a \right) \left( D_\mu \phi^a \right) - V_{oct}( \phi^a \phi^a )
 \label{lagrangeano}
\end{eqnarray}
where $D_\mu = \partial_\mu + i g_M T^a M^a_\mu$  is the covariant derivative,
$T^a$ stands for the generators of $SU(3)_Q$ group and
$V_{oct}$ is the potential energy associated with $\phi^a$. 
Note that the second term in (\ref{lagrangeano}) includes a sum over all families of fermions.
In $\mathcal{L}$ the terms associated with the SM for each sector $s = 1, \cdots, N_Q$
and those associated with the quantization of the theory are omitted. 

The kinetic term associated with the scalar field accommodates a mass term for the Mulato field. The gauge field
mass term is associated with the operator
\begin{equation}
  \frac{1}{2} \, g^2_M \, \phi^c (T^a T^b)_{cd} \phi^d M^a_\mu M^{b \, \mu} \, .
\end{equation}
The scalar field cannot acquire a vacuum expectation value without breaking gauge invariance. However,
to generate a mass for the Mulato it is sufficient to assume a non-vanishing boson condensate
$\langle \phi ^a \phi^b \rangle$. The origin of this condensate can be associated with local fluctuations of
the scalar field. 

If the dynamics of the scalar field is such that
\begin{equation}
\langle \phi ^a \rangle = 0 \qquad\mbox{ and }\qquad \langle \phi^a \phi^b \rangle = v^2 \delta^{ab} \, ,
\label{condensado}
\end{equation}
given that for the adjoint representation $\mbox{tr} (  T^a T^b )= 3 \, \delta^{ab}$, it follows that
the square of the Mulato mass reads
\begin{equation}
 M^2  = 3 \, g^2_M   v^2 \, .
 \label{WIMG_mass}
\end{equation} 
Note that $v^2$ and, therefore,  the Mulato mass are gauge invariant. The proof of gauge invariance follows 
directly from the transformations properties of $\phi^a$ \cite{Orlando}. 

\section{Big Bang Nucleosynthesis and Baryon Asymmetries} 

The gauge model summarized in (\ref{lagrangeano}) has new relativistic degrees of freedom that can
increase the expansion rate of the early Universe \cite{Berezhiani1996} and affect  
the BBN \cite{Hoyle}.
After inflation, the temperature for the thermal baths associated with each particle species
is not necessarily the same \cite{Berezhiani1996}. It depends on the various possible reactions enabling equilibria
and on the Universe thermal history. Let us start discussing the simplest possible picture where all the
dark sectors have the same temperature, different from the ordinary matter thermal bath, i.e. we are assuming 
that asymmetric reheating takes place after inflation as in \cite{Kolb,Berezhiani2001,Ciarcelluti}.

The number of possible new particles contributing to the radiation density during the BBN epoch are
constrained by the $^4$He primordial abundance and the baryon-to-photon ratio
$\eta = n_b/n_\gamma$, where $n_b$ is the baryon density and $n_\gamma$ the photon density in the Universe
\cite{BBNLimit}.
For a radiation dominated Universe at very high temperatures, neglecting the particles masses,
the energy and entropy densities are given by \cite{pdg2010}
\begin{equation}
 \rho(T)=\frac{\pi^{2}}{30} \, g_*(T) \,T^4 \quad\mbox{and}\quad
 s(T)=\frac{2\pi^{2}}{45} \, g_s(T)\, T^3 ,
 \label{energy_entropy}
\end{equation}
where
\begin{equation}
g_*(T)=\sum_B g_B \left(\frac{T_{B}}{T}\right)^4 + \frac{7}{8} \sum_F g_F \left(\frac{T_{F}}{T}\right)^4
\end{equation}
and
\begin{equation}
g_s(T)=\sum_B g_B \left(\frac{T_{B}}{T}\right)^3 + \frac{7}{8} \sum_F g_F \left(\frac{T_{F}}{T}\right)^3,
\end{equation}
are the effective number of degrees of freedom during nucleosynthesis, $g_{B(F)}$ is the number of degrees of
freedom of the boson (fermion) species $B(F)$, $T_{B(F)}$ is the temperature of the thermal bath of species 
$B(F)$ and $T$ the temperature of the photon thermal bath.

In our case, we consider that the ordinary and dark sectors are decoupled, just after reheating, with different
temperatures: $T$ for ordinary matter and $T'$ for the dark sectors. For the dark sectors,
the energy $\rho'(T')$ and entropy $s'(T')$ densities are given as in (\ref{energy_entropy})
after replacing $g_*(T) \rightarrow g'_*(T')$ and $g_s(T) \rightarrow g'_s(T')$, i.e.
the effective number of degrees of freedom in the dark sector, and replacing $T$ by $T'$.
The entropy in each sector is separately conserved during the Universe evolution,
which leads that $x=(s'/s)^{1/3}$ is time independent. Assuming the same relativistic particle
content for each sector of the modern universe, one has  $g_s(T_0) = g_s^\prime(T'_0)$ and
it follows that  $x=T'/T$.

For a radiation dominated era, the Friedman equation is
\begin{equation}
H(t)=\sqrt{\left(8\pi/3 c^2\right) \, G_{N} \, \bar{\rho}},
\end{equation}
where the total energy density is given by $\bar{\rho} = \rho\, + \, N_{DM} \, \rho'$, where
$N_{DM} = N_Q - 1$ is the number of dark sectors. From the expression for $\rho'$,
it follows
\begin{equation}
H(t)=1.66 \, \sqrt{\bar{g}_{*}(T)} \, \frac{T^2}{M_{Pl}},
\end{equation}
where
\begin{equation}
\bar{g}_{*}(T) = g_{*} (T) \left( 1+ N_{DM} \, a \, x^4 \right),
\end{equation}
and $M_{Pl}$ is the Planck mass.
The parameter $a = \left(g'_{*}/g_{*}\right) \left(g_{s}/g'_{s}\right)^{4/3}\sim 1$, unless $T'/T$ is very
small \cite{Berezhiani1996}. At the nucleosynthesis temperature scale of about 1 MeV, the
relativistic degrees of freedom (photons, electrons, positrons and neutrinos)
are in a quasi-equilibrium state
and $g_{*}(T)|_{T=1MeV} = 10.75$.
The extra dark particles
change $g_{*}$ to $\bar{g}_{*} = g_{*} \left(1 + N_{DM} \, x^4 \right)$. The bounds due to the relative
abundances of the light element isotopes ($^{4}$He, $^{3}$He, D and $^{7}$Li) are usually
written in terms of the equivalent number of massless neutrinos during nucleosynthesis:
$3.46 <  N_{\nu} < 5.2$ \cite{WMAP-cosm}.
The extra degrees of freedom introduced by the dark sectors lead to
$\Delta g_{*}=\bar{g}_{*}-g_{*} = 1.75 \, \Delta N_\nu < 3.85$, where $\Delta N_\nu$ is the variation
in equivalent number of neutrinos, and  $T' / T < 0.78 / N^{1/4}_{DM}$ to reconcile the gauge model
with the BBN data.
If $N_{DM} = 5$, as required by to explain the observed ratio $\Omega_{DM} / \Omega_b$, then
$T' /T < 0.52$. In conclusion, the asymmetric reheating mechanism leads always to dark universes
which are colder than our one universe.

The baryon asymmetry is parameterized by the baryon-to-photon ratio $\eta$.
The density number of photons $n_{\gamma}$ is proportional to $T^{3}$ and, therefore,
one can write the density number of dark-photons as $n'_{\gamma}=x^3 n_{\gamma}$. The
ratio of dark-baryons to ordinary-baryons is given by $\beta=\Omega'_{B}/\Omega_{B} = x^3 \eta'/\eta$
\cite{Berezhiani2001}. The bounds from the BBN on $x = T' / T$ imply that the baryon asymmetry in the
dark sector is greater than in the ordinary one. Indeed, using the upper bound $x\sim 0.78/N^{1/4}_{DM}$ 
and assuming that each sector contributes equally to the Universe's energy
density $\beta\sim 1$, we obtain $\eta' \sim 2.1 \, N^{3/4}_{DM}\eta$. For the special where
$N_{DM} = 5$ it follows that $\eta'\sim 7\eta$. Asymmetric Dark Matter models, see e.g.
\cite{ADM1,ADM2,ADM3,ADM4}, give similar results for the baryon asymmetry.

In principle, the presence of mirror baryon dark matter (MBDM) could give some effect on the CMB power 
spectrum. The reason is that the acoustic oscillations of MBDM could be transmitted to the ordinary baryons. 
In Ref. \cite{Berezhiani2001} this effect was analyzed and their conclusion is that to obtain
an observable effect in CMB data it is necessary to have a ratio of temperatures $T'/T\geq 0.6$. 
This bound combined with the BBN analysis provides a lower bound for the number of dark sectors:
$0.35 < N_{DM}$.

Galaxy dynamics provide further constraints on DM, see e.g.
\cite{Mohapatra2002,Ackerman2009}.
In the gauge model there is no direct coupling between the photon and its dark brothers.
Further, it is assumed that the different sectors behave as the ordinary matter family.
It seems natural that the galaxy halos are neutral relative to the $U(1)$'s within each
sector.
The observed dark matter halos suggest that DM are effectively
collisionless and demand an upper bound in the cross section of DM-DM interactions
\cite{Yoshida2000,Wandelt2000,Escude2000,Dave2001}.
The $T'/T$ bound estimated from BBN complies with such a statement.
A typical cross section is given by
$\sigma \approx ( g^2 T / \Lambda^2)^2$, where $g$ is the interaction coupling constant,
$T$ is the temperature and $\Lambda$ a typical mass scale of the interaction.
If the dark sectors are copies of the ordinary matter sector, i.e. $g$ and $\Lambda$ are of the same
order of magnitude, one can write $\sigma'/\sigma = (T'/T)^2$, where
$\sigma'$ ($\sigma$) is the cross section for the dark (ordinary) family. The temperature
bound from BBN 
implies that $\sigma' / \sigma < 0.61/\sqrt{N_{DM}}$
and, as long as $T' / T$ is sufficiently small, DM becomes effectively
collisionless.
This difference in the temperature seems to suggest
that the dark sectors are not exact copies of the SM sector. Dark Universes are very cold. 

\section{Electron Screening during BBN}

Modeling the BBN and  stellar evolution  requires that one includes  the information on nuclear reaction rates $\langle \sigma v\rangle$ in reaction network calculations, where $\sigma$ is the nuclear fusion cross sections and $v$ is the relative velocity between the participant nuclides. Whereas $v$ is well described by a Maxwell-Boltzmann velocity distribution for a given temperature $T$, the cross section $\sigma$ is taken from laboratory experiments on earth, some of which are not as well known as desired \cite{Nol,Serpico:2004gx,Cyburt:2001,Cyburt:2002,Cyburt:2004cq,BBH97}. 
Using the Debye-H\"uckel model, Salpeter \cite{Sal54} showed that stellar electron screening enhances cross sections, yielding an enhancement factor.
The  Debye-Hueckel model used by Salpeter yields a screened Coulomb potential, valid when $\langle V\rangle \ll kT$ (weak screening),
which depends on the ratio of the Coulomb potential
at the Debye radius $R_D$ \cite{Sal54}. 

\begin{figure}[t]
{\includegraphics[width=12.cm]{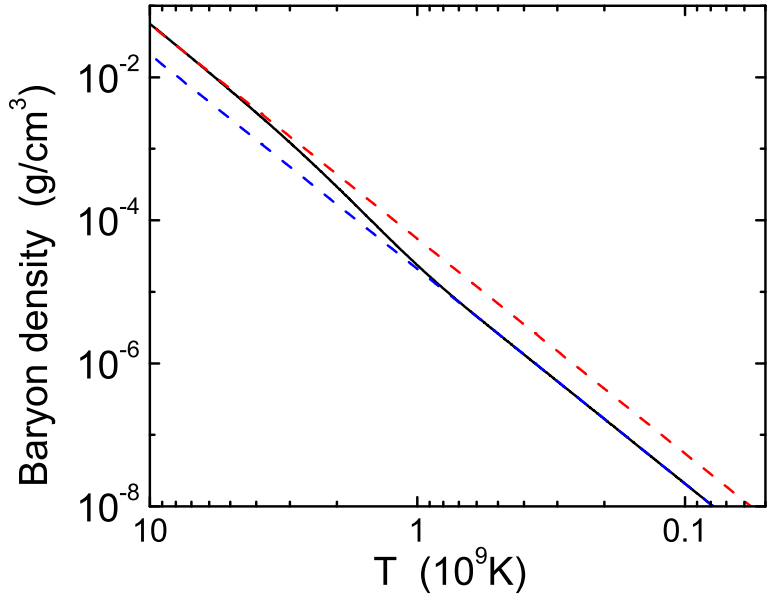}}
\caption{
Baryon density (solid curve) during the early universe as a function of the temperature in units of billion degrees Kelvin, $T_9$.   The dashed curves are obtained from Eq. \eqref{rhoT} with $h\sim 2.1\times 10^{-5}$ and $h\sim 5.7\times 10^{-5}$, respectively.  
}\label{fig1}
\end{figure}

Corrections 
to the Salpeter formula are expected at some level.  
Dynamic corrections were first discussed 
by \cite{Mit77} and later studied by \cite{CSK88}.
Subsequent work showed that Salpeter's formula would be valid
independent of the Gamow energy due to the nearly precise thermodynamic
equilibrium of the solar plasma \cite{BS97,Gru98,GB98}. 
Later, a number of contradictions were pointed out in
investigations claiming larger corrections, and a field theoretic approach
was shown to  lead to the expectation
of only small ($\sim$ 4\%) corrections to the standard formula, for solar conditions \cite{BBGS02}. 

A good measure of the screening effect is given by the screening parameter given  by $\Gamma = Z_1Z_2e^2/\langle r \rangle kT$, where $\langle r \rangle =n^{-1/3}$.  In the core of the sun densities are of the order of $\rho \sim 150$ g/cm$^3$ with temperatures of $T\sim 1.5 \times 10^7$ K.  For pp reactions in the sun, we thus get $\Gamma \sim 1.06$ which validates the weak screening approximation, but for p$^7$Be reactions one gets $\Gamma_{{\rm p}^7{\rm Be}}\sim 1.5$, which is one of the reasons to support modifications of the Salpeter formula. Also, in the sun the number of ions within a sphere of radius $R_D$ (Debye sphere) is of the order of $N\sim 4$. As the Debye-Hueckel approximation is based on the mean field approximation, i.e., for $N=n(4\pi R^3/3) \gg 1$, deviations from the Salpeter approximation are justifiable. 

The electron density during the early universe varies strongly with the temperature as seen in figure \ref{fig1}, where $T_9$ is the temperature in units of $10^9$ K ($T_9$). This can compared with the electron number density at the center of the sun, $n_e^{sun}\sim 10^{26}$/cm$^3$.  The figure shows that, at typical temperatures $T_9 \sim 0.1-1$ during the BBN the universe had electron densities which are much larger that the electron density in the sun.  However, in contrast to the sun, the baryon density in the early universe is much smaller than the electron density. The large electron density  is due to the $e^+e^-$ production by the abundant photons during the BBN. 

The baryonic density is best seen in
figure \ref{fig1}. It varies as
\begin{equation}
\rho_b \simeq h T_9^3, 
\label{rhoT}
\end{equation} 
where $h$ is the baryon density parameter \cite{SW77}. It can be calculated  by using Eq. (3.11) of Ref. \cite{SW77} and the baryon-to-photon ratio $\eta=6.19\times 10^{-10}$  at the BBN epoch (from WMAP data \cite{Kom10}). Around $T_9\sim 2$ there is a change of the value of $h$ from $h\sim 2.1\times 10^{-5}$ to $h\sim 5.7\times 10^{-5}$.  Eq. \eqref{rhoT} with the two values of $h$ are shown as dashed lines in figure \ref{fig1}, obtained in Ref. \cite{WBB11}.

\begin{figure}[t]
{{\includegraphics[width=12.cm]{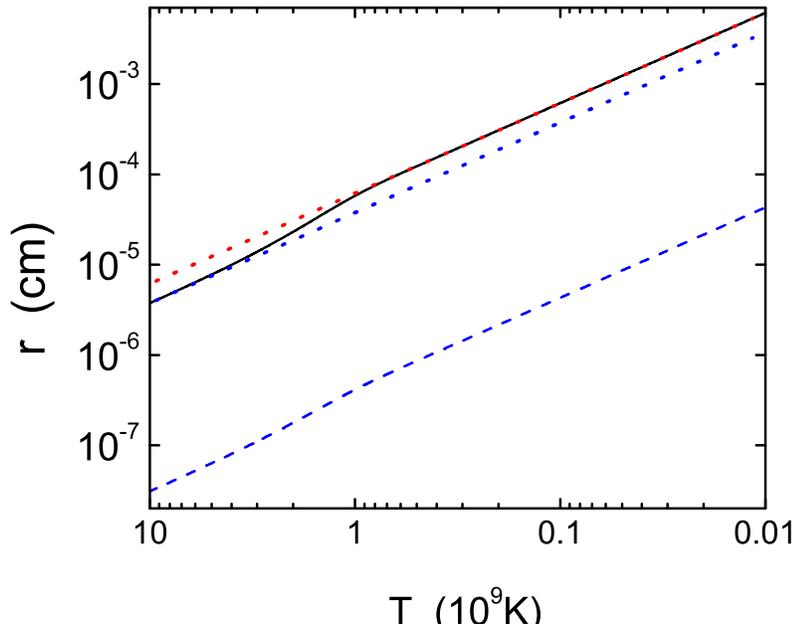}}}
\caption{
Debye radius during the BBN as a function of the temperature in units of billion degrees Kelvin (solid line). The dotted lines are the approximation given by Eq. \eqref{rd} with  $R_D^{(0)}\sim 6.1\times 10^{-5}$ cm and $R_D^{(0)}\sim 3.7\times 10^{-5}$ cm, respectively. The inter-ion distance is shown by the isolated dashed-line. }\label{fig2}
\end{figure}

It is also worthwhile to calculate the Debye radius as a function of the temperature. This is shown in figure  \ref{fig2}. The accompanying dashed lines correspond to the approximation of Eq. \eqref{rhoT}, with $h\sim 2.1\times 10^{-5}$ and $h\sim 5.7\times 10^{-5}$. This leads to two straight lines in a logarithmic plot of 
\begin{equation}
R_D = {R_D^{(0)}  T_9^{-1}},
\label{rd}
\end{equation}
with $R_D^{(0)}\sim 6.1\times 10^{-5}$ cm and $R_D^{(0)}\sim 3.7\times 10^{-5}$ cm, respectively. In figure \ref{fig2} we also show the inter-ion distance by the lower dashed line. It is clear that the number of ions inside the Debye sphere is  at least of the order $10^3$, which would justify the mean field approximation for the ions.
In contrast to protons, electrons and positrons are mostly relativistic and their chaotic motion will probably average out the effect of screening around the ions. But because the number density of electrons is large, an appreciable fraction of them still carry velocities comparable to those of the ions. 

Using a standard numerical computation of the BBN we have shown that electron screening cannot be a source of measurable changes in the elemental abundance. This is verified by artificially increasing the screening obtained by traditional models \cite{Sal54}. We back our numerical results with very simple and transparent estimates. This is also substantiated by the mean-field calculations of screening due to the more abundant free $e^+e^-$ pairs published in Ref. \cite{Itoh97}. They conclude that screening due to free pairs might yield a 0.1\% change on the BBN abundances.   But even if mean field models for electron screening were not reliable under certain conditions, which we have discussed thoroughly in the text, it is extremely unlikely that electron screening might have any influence on the predictions of the standard Big Bang nucleosynthesis.

\section{BBN with Non-Extensive Statistics}

An increasing number of experiments, theoretical developments, have challenged the  Boltzmann-Gibbs description of statistical mechanics. It seems that the Boltzmann-Gibbs is not adequate for systems with long range interactions, and with memory effects. Therefore, it was unavoidable that new approaches for the Boltzmann-Gibbs formalism were proposed. Nowadays, a very popular approach is based on the proposal by Tsallis \cite{Ts88}, who replace the traditional entropy by the following one:
\begin{equation}
S_q=k_B {1-\sum_i p^q_i \over q-1},
\end{equation}
where $p_i$ is the probability to find the system in the microstate $i$, $q$ is a real number. For $q=1$, $S_q= S_{BG}$, and $S_q$ is a natural generalization of the Boltzmann-Gibbs entropy.

Based on the successes of the big bang model, it is fair to assume that it can set strong constraints on the limits of the parameter $q$ used in  a non-extensive statistics description of the Maxwell-Boltzmann velocity distribution. In the literature, attempts to solve the lithium problem has assumed all sorts of ``new physics" \cite{WBB11,Fie11}. The work presented in Ref. \cite{BFH12} adds to the list of new attempts, although our results imply a much wider impact on BBN as expected for the solution of the lithium problem. If the Tsallis statistics appropriately describes the deviations of tails of statistical distributions, then the BNN would effectively probe such tails.  The Gamow window contains a small fraction of the total area under the velocity distribution.  Thus, only a few particles in the tail of the distributions contribute to the fusion rates.   In fact,  the possibility of a deviation of the Maxwellian distribution and implications of the modification of the Maxwellian distribution tail for nuclear burning in stars have already been explored in the past (see \cite{BFH12} and references therein).  In Ref. \cite{BFH12} it was shown that a strong deviation from $q=1$ is very unlikely for the BNN predictions, based on comparison with observations. Moreover, if $q$ deviates from the unity value, the lithium problem gets even worse \cite{BFH12}. 


\begin{acknowledgements}
{\small The authors acknowledge financial support from the Brazilian
agencies FAPESP (Funda\c c\~ao de Amparo \`a Pesquisa do Estado de
S\~ao Paulo) and CNPq (Conselho Nacional de Desenvolvimento
Cient\'ifico e Tecnol\'ogico) and the US Department of Energy
Grants DE-FG02-08ER41533, DE-SC0004971. OO acknowledges financial support from FCT under
contract PTDC/\-FIS/100968/2008.}
\end{acknowledgements}


\end{document}